\documentclass[journal]{IEEEtran}
\usepackage{cite}
\usepackage{amsmath,amssymb,amsfonts}
\usepackage{graphicx}
\usepackage{textcomp}
\usepackage{dcolumn}
\usepackage{bm}
\usepackage{subfig} 
\usepackage{xcolor}

\ifCLASSINFOpdf
\else
\fi

\begin{document}

\onecolumn
\newpage
\thispagestyle{empty}

\textbf{Copyright:}
\copyright 2020 IEEE. Personal use of this material is permitted.  Permission from IEEE must be obtained for all other uses, in any current or future media, including reprinting/republishing this material for advertising or promotional purposes, creating new collective works, for resale or redistribution to servers or lists, or reuse of any copyrighted component of this work in other works.\\

\textbf{Disclaimer:} This work has been published in \textit{IEEE Transactions on Microwave Theory and Techniques}. \\

Citation information: DOI 10.1109/TMTT.2020.3011004

\newpage

\twocolumn

%
\title{Glide-Symmetric Metallic Structures with Elliptical Holes for Lens Compression}
%
%
%

\author{Antonio Alex-Amor, Fatemeh Ghasemifard, \IEEEmembership{Graduate Student Member,~IEEE,} Guido Valerio,~\IEEEmembership{Senior Member,~IEEE}, Mahsa Ebrahimpouri, \IEEEmembership{Student Member,~IEEE,}
        Pablo Padilla, Jos\'{e} M. Fern\'{a}ndez-Gonz\'{a}lez, \IEEEmembership{Senior Member,~IEEE}, and Oscar Quevedo-Teruel,~\IEEEmembership{Senior Member,~IEEE}
 \thanks{Manuscript received X, 2019; revised X, 2019. This work was partially funded by the  Ministerio de Ciencia Innovaci\'{o}n y Universidades under the project TIN2016-75097-P, with European Union FEDER funds under the project FUTURERADIO  ``Radio  systems  and  technologies  for  high  capacity  terrestrial  and  satellite  communications  in  an  hyperconnected world" (project number TEC2017-85529-C3-1-R), by the French National Research Agency Grant Number ANR-16-CE24-0030, by the Vinnova project High-5 (2018-01522) under the Strategic Programme on Smart Electronic Systems, and by the Stiftelsen \AA forsk project H-Materials (18-302).}
\thanks{A. Alex-Amor is with the Departamento de Lenguajes y Ciencias de la Computaci\'{o}n, Universidad de M\'{a}laga, 29071 M\'{a}laga,  Spain.}
\thanks{A. Alex-Amor and J. M. Fern\'{a}ndez-Gonz\'{a}lez are with the Information Processing and Telecommunications Center, Universidad Polit\'{e}cnica de Madrid, 28040 Madrid, Spain (e-mail: aalex@gr.ssr.upm.es; jmfdez@gr.ssr.upm.es)  }
\thanks{F. Ghasemifard, M. Ebrahimpouri and O. Quevedo-Teruel are with the Department of Electromagnetic Engineering, School of Electrical Engineering and Computer
	Science, KTH Royal Institute of Technology, SE-100 44 Stockholm, Sweden
	(e-mail: fatemehg@kth.se; mahsaeh@kth.se; oscarqt@kth.se)  }
\thanks{G. Valerio is with UR2, Laboratoire d'\'{E}lectronique and \'{E}lectromagn\'{e}tisme, Sorbonne Universit\'{e}, F-75005 Paris, France (e-mail: guido.valerio@sorbonne-universite.fr)  }
\thanks{P. Padilla is with the Departamento de Teor\'{i}a de la Se\~{n}al, Telem\'{a}tica y Comunicaciones, Universidad de Granada, 18071 Granada, Spain (email: pablopadilla@ugr.es)   }
}

%
%

\markboth{IEEE Transactions on Microwave Theory and Techniques,~Vol.~X, No.~X, ~2020}%
{Alex-Amor \MakeLowercase{\textit{et al.}}: Glide-Symmetric Metallic Structures with Elliptical Holes for Lens Compression}
%



\maketitle

\begin{abstract}
In this paper, we study the wave propagation in a metallic parallel-plate structure with glide-symmetric elliptical holes. To perform this study, we derived a mode-matching technique based on the generalized Floquet theorem for glide-symmetric structures. This mode-matching technique benefits from a lower computational cost since it takes advantage of the glide symmetry in the structure. It also provides physical insight on the specific properties of Floquet modes propagating in these specific structures. With our analysis, we demonstrate that glide-symmetric structures with periodic elliptical holes exhibit an anisotropic refractive index over a wide range of frequencies. The equivalent refractive index can be controlled by tuning the dimensions of the holes.
Finally, by combining the anisotropy related to the elliptical holes and transformation optics, a Maxwell fish-eye lens with a 33.33\% size compression is designed. This lens operates in a wideband frequency range from 2.5 GHz to 10 GHz.
\end{abstract}


%
\IEEEpeerreviewmaketitle

\section{Introduction}
%
%
%
%

\IEEEPARstart{M}{etamaterials} are artificial materials formed by subwavelength periodic structures that exhibit extraordinary macroscopic properties \cite{nature, metamaterials,metamaterials2, book_metamaterial}. Metasurfaces \cite{metarsurfaces0, metasurfaces2, silicon}, the two-dimensional version of metamaterials, opened the possibility of manufacturing low-profile and cost-effective devices \cite{metasurfaces1}. Metasurfaces have been used for controlling (allowing or suppressing) the propagation of electromagnetic waves in a given direction \cite{kildal1,kildal2}, to produce electromagnetic cloaking \cite{cloak1,cloak2}, wavefront transformation \cite{wavefront1,wavefront2}, and graded-index lenses \cite{gradex_index_lens1, mahsa2, lens_oscar}. 

Recent  studies have demonstrated the distinctive dispersive properties related with the presence of higher symmetries in periodic structures. For example, introducing  glide and twist symmetries in periodic structures can increase  their equivalent refractive index \cite{fatemeh,nature2}. This feature has been used for antenna miniaturization \cite{symmetry}, mutual coupling reduction \cite{angel_iceea} or to produce  a compact and low-loss phase shifter at millimeter waves \cite{access_angel}. Furthermore, with the use of higher symmetries, the stopband between the first and the second modes of conventional  periodic structures is suppressed \cite{mode_matching1,pablo,adri} and the frequency dispersion of the first propagating modes is reduced \cite{dahlberg,camacho,qiao,fatemeh}. The latter can be used for realizing fully-metallic low-loss wideband lenses for 5G communication systems \cite{lens,lens_oscar}. Additionally, it has been recently proved that breaking symmetries  can be used for filtering purposes \cite{pablo2}.

Previous works have studied in detail the implementation of \emph{isotropic} holey metasurfaces \cite{garcia_vidal,garcia_deabajo,antonio_triangular}, composed of squared and circular holes. In \cite{antonio_triangular}, a comparison between holey metasurfaces composed of square, circular and triangular holes is presented. It is shown that square and triangular holes possess higher equivalent refractive indexes and wider stopbands compared to the configuration that uses circular holes. On the other hand, the frequency dispersion of the structure with circular holes is normally lower compared to triangular and square holes. However, some specific lenses require a certain level of \emph{anisotropy}, for example, optically transformed lenses \cite{TO1,TO2,mahsa}. In this case, rectangular holes can provide the necessary anisotropy \cite{plasmonic_surfaces, guido}. However, rectangular holes are difficult to manufacture with milling techniques. Alternatively, holey structures with elliptical holes are a practical solution to implement anisotropy in parallel-plate structures. In order to tune the equivalent refractive index, a change of the ellipticity and depth of the holes can be applied, similarly to circular holes in isotropic implementations \cite{fateme2h}.

 Glide-symmetric periodic structures require specific modelling tools due to the strong interaction between the layers  \cite{bloch, mesa}. Even though circuit approaches \cite{qiao} and homogenized impedance models \cite{kildal3} are fast tools to characterize the dispersion properties of a great variety of periodic structures, these methods are difficult to implement for a glide-symmetric configuration due to the strong coupling between the metallic layers. Furthermore,  while integral-equation formulations can handle multiple types of geometries \cite{integral_equation1,integral_equation2,integral_equation3}, they do not give physical insight on the operation of the structure due to their purely numerical approach. In this paper, we apply a general mode-matching technique to study the dispersion features and the anisotropy related to glide-symmetric metallic structures with periodic elliptic holes. Our results confirm the wideband nature of glide-symmetric structures. 
	 
	  
	 This paper  is organized as follows. Section II presents the general mode-matching formulation applied to the elliptical holey structure. The fields inside the holes are described by means of Mathieu functions, and they are subsequently used to obtain the dispersion equation. Section III presents the dispersion analysis for all the geometric parameters of the structure. Furthermore, a study on the anisotropic behavior of the equivalent refractive index is also carried out. Section IV presents the design of a compressed Maxwell fish-eye lens made of glide-symmetric periodic elliptical holes. Finally, conclusions are drawn in Section V.


\section{Formulation}
In this section, we propose a mode-matching technique to analyze the 2D periodic holey structure composed of elliptical holes shown in Fig. \ref{unit_cells}(a). Fig. \ref{unit_cells}(b) shows a top view of the structure when the metallic plate is removed. Mat and bright red colors in Fig. \ref{unit_cells}(b) relate the elliptical holes of the bottom and top layers. Fig. \ref{unit_cells}(c)  depicts the glide-symmetric version of the holey metasurface, and Fig. \ref{unit_cells}(d) the non-glide-symmetric version.  Both structures are periodic, with period $d$, along the $x$- and $y$-directions and are bounded along the $z$-direction. The gap region is filled with a dielectric of relative permittivity $\varepsilon_r$ and the elliptical holes are filled with a dielectric of relative permittivity $\varepsilon_{r\mathrm{hole}}$. The $z = 0$ plane is chosen to be on the top of the  non-glide-symmetric unit cell, and in the center of the gap between plates in the glide-symmetric case. The gap between the plates is $g$ and $g/2$ in the glide-symmetric and non-glide-symmetric structures, respectively. 


\subsection{Mode-matching Formulation}

\begin{figure}[t]
	\vspace*{-0.3cm}
	\centering
	\subfloat[\hspace*{-0.8cm}]{
		\hspace*{-0.5cm}
		\label{holey}
		\includegraphics[width=0.25\textwidth]{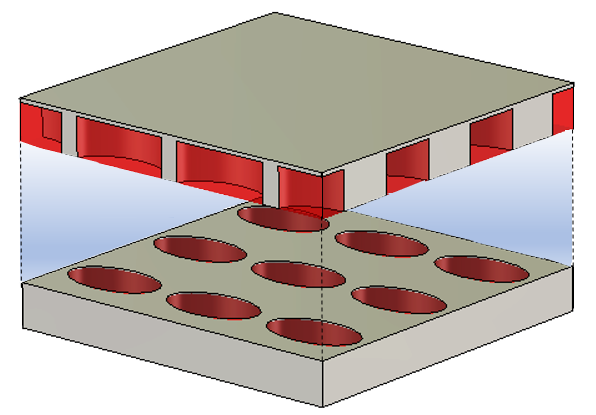}}
	\subfloat[\hspace*{-0.4cm}]{
		\label{topview_holey}
		\includegraphics[width=0.19\textwidth]{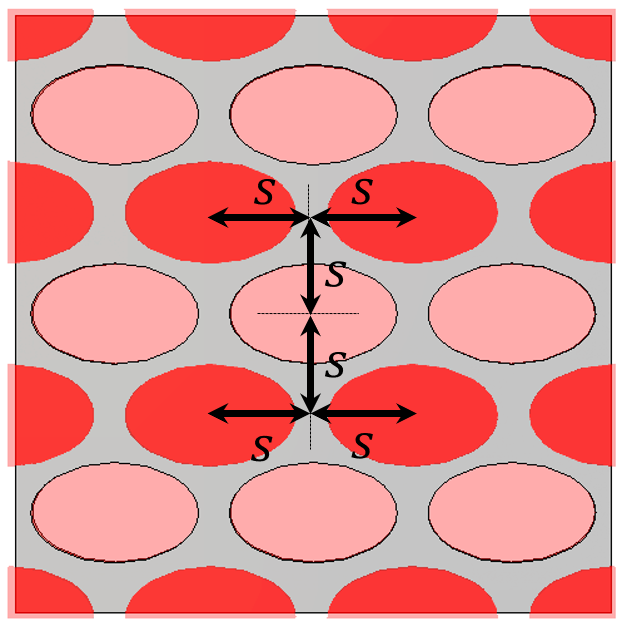}}\\
	\vspace*{-0.3cm}
	\subfloat[\hspace*{-0.8cm}]{
		\label{glide}
		\includegraphics[width=0.23\textwidth]{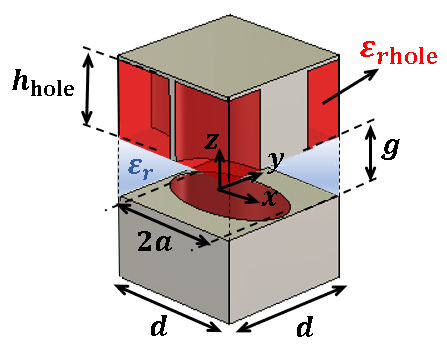}}
	\subfloat[\hspace*{-0.5cm}]{
		\hspace*{0.4cm}
		\label{nonglide}
		\includegraphics[width=0.18\textwidth]{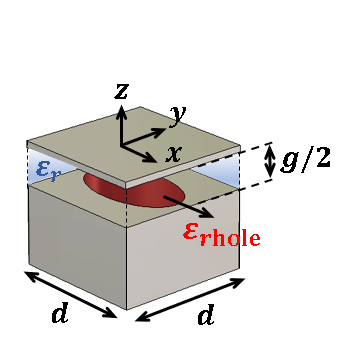}}\\
	\subfloat[]{
		\label{15V}
		\includegraphics[width=0.22\textwidth]{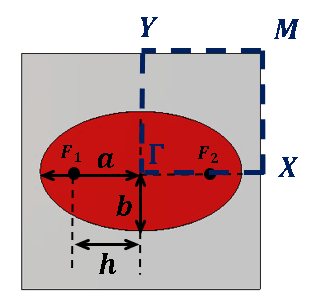}}\\
	\caption{\small (a) 2D periodic holey structure and (b) a top view of it. Unit cells of the (c) glide-symmetric ($s=d/2$) and (d) non-glide-symmetric configurations. (e) Cross section of an elliptical hole .}
	\label{unit_cells}
\end{figure}

In this section, first, we briefly present the mode-matching formulation for this problem in a suitable coordinate system. An elliptic cylindrical coordinate system  is used due to the geometry of the hole. As stated in Appendix A, the elliptic cylindrical coordinate system is defined by three coordinates: radial $\xi$, angular $\eta$, and longitudinal $z$. Using this coordinate system, the tangential fields in the surface of the hole can be expressed as 
\begin{equation} \label{modal_e}
	\textbf{E}_t^{\mathrm{WG}} (\xi,\eta,z=-g/2)= \sum_{i=1}^N r_i^- C_i \mathbf{\Phi_i} (\xi,\eta) 
\end{equation}
\begin{equation} \label{modal_h}
\textbf{H}_t^{\mathrm{WG}} (\xi,\eta,z=-g/2)= \sum_{i=1}^N r_i^+ Y_i C_i \left[\mathbf{\hat{z}} \times \mathbf{\Phi_i} (\xi,\eta) \right]
\end{equation}%
where $C_i$ is the coefficient of the $i$-th mode, $\mathbf{\Phi_i} (\xi,\eta) $ is the modal function that represents the $i$-th mode inside the hole, and $Y_i$ is the wave admitance of the $i$-th mode. The elliptical hole can be regarded as a short-circuited elliptical waveguide. Thus, the coefficients $r_i^{\pm}$ in \eqref{modal_e} and \eqref{modal_h} are calculated as
\begin{equation}
	r_i^{\pm}= 1 \pm e^{-j2k_{zi}h_{\mathrm{hole}}}.
\end{equation}
 The term $k_{zi}=\sqrt{\varepsilon_{r\mathrm{hole}}k_0^2-k_{ti}^2}$ is the wavenumber  in the longitudinal direction of the elliptical waveguide, $\varepsilon_{r\mathrm{hole}}$ is the relative dielectric constant inside the hole, $k_{ti}$ is the transversal wavenumber and $h_{\mathrm{hole}}$ is the hole depth in Fig. \ref{unit_cells}. 

On the other hand, the tangential fields in the gap region can be expressed as a series of Floquet harmonics:
\begin{equation} \label{Floquet_e}
\textbf{E}_t^{\mathrm{GAP}}= \frac{1}{d^2} \sum_{p} \sum_{s} e^{-j(k_{x,p}x+k_{y,s}y)}\, \mathbf{\tilde{e}}_{t,ps}^{\mathrm{GAP}}(z)
\end{equation}
\begin{equation} \label{Floquet_h}
\textbf{H}_t^{\mathrm{GAP}}= \frac{1}{d^2} \sum_{p} \sum_{s} e^{-j(k_{x,p}x+k_{y,s}y)}\, \mathbf{\tilde{h}}_{t,ps}^{\mathrm{GAP}}(z)
\end{equation}
where $k_{x,p}=k_{x,0}+\frac{2 \pi p}{d}$, $k_{y,s}=k_{y,0}+\frac{2 \pi s}{d}$, $p$ and $s$ are the integer values specifying the order of the Floquet harmonics, and $\mathbf{\tilde{e}}_{t,ps}^{\mathrm{GAP}}(z)$, $\mathbf{\tilde{h}}_{t,ps}^{\mathrm{GAP}}(z)$ are the field amplitudes of each transversal electric and magnetic Floquet harmonic. Both field-amplitude terms are expressed as a sum of sine and cosine functions \cite{fateme2h}.

By applying the boundary conditions (continuity of electric and magnetic tangential fields at $z=-g/2$ and $z=g/2$ for the glide case, and at $z=-g/2$ and $z=0$ for the non-glide case), the following linear system of equations is obtained  \cite{fateme2h,fatemeh_tesis}:
  \begin{equation} \label{linear_system}
  	  \sum_{i=1}^N C_i \, \alpha_{ri} =0, \qquad r=1, ..., N
  \end{equation}
where $N$ is the number of modal functions considered in the elliptical hole and
\begin{equation} \label{eigen}
\alpha_{ri}=jk_0 \eta_0 d^2 Y_i I_{ri} + \frac{r_i^-}{r_i^+} \sum_{p} \sum_{s} \widetilde{f}_{ps} (k_{z,ps}) \beta_{ri} (\mathbf{k}_{ps}).
\end{equation}
By setting the determinant of the coefficient matrix $\alpha_{ri}$ to zero, the dispersion equation can be obtained. In \eqref{eigen}, $I_{ri}$ is the inner product of the $r$-th and $i$-th modal functions. The parameter $\beta_{ri} (\mathbf{k}_{ps})$ is expressed as
\begin{multline} \label{beta}
	\beta_{ri} (\mathbf{k}_{ps})= \beta_{ri} (k_{x,p}, k_{y,s}, k_{z,ps})\\= 
	\frac{k_0^2-k_{y,s}^2}{k_{z,ps}}\, \widetilde{\phi}^x_i \widetilde{\phi}^{x*}_r + 
	\frac{k_{x,p} k_{y,s}}{k_{z,ps}}\, \widetilde{\phi}^y_i \widetilde{\phi}^{x*}_r \\+
	\frac{k_0^2-k_{x,p}^2}{k_{z,ps}}\, \widetilde{\phi}^y_i \widetilde{\phi}^{y*}_r +
	\frac{k_{x,p} k_{y,s}}{k_{z,ps}}\, \widetilde{\phi}^x_i \widetilde{\phi}^{y*}_r 
\end{multline}
where $k_{z,ps}=\sqrt{\varepsilon_rk_0^2-k_{x,p}^2-k_{y,s}^2}$ is the vertical wavenumber of the $(p,s)$-th harmonic, $\varepsilon_r$ is the relative dielectric constant in the gap region, $\widetilde{\phi}^x_i$ and $\widetilde{\phi}^y_i$ are the $x$ and $y$-components of the Fourier transform of the  $i$-th modal function,   and the symbol (*) denotes the complex conjugate. Finally, the vertical spectral function $\widetilde{f}_{ps}$ in \eqref{eigen} distinguishes between the glide-symmetric and non-glide-symmetric cases. In the non-glide-symmetric case, it is
\begin{equation} \label{spectral_nonglide}
\widetilde{f}_{ps}= \cot \left(k_{z,ps}\, g/2 \right)
\end{equation}
and in the glide-symmetric case, it is
\begin{equation} \label{spectral_glide}
\widetilde{f}_{ps}= 
\begin{cases}
-\tan \left(k_{z,ps}\, g/2 \right) \qquad p+s \, \, \, \textrm{even}\\
\, \, \, \, \,\cot \left(k_{z,ps}\, g/2 \right) \qquad \,  p+s \, \, \, \textrm{odd}\\
\end{cases}
\end{equation}

\subsection{Numerical Implementation}

Elliptical holes can be seen as short-circuited elliptical waveguides. The foundations of elliptical waveguides were first established in \cite{elliptic_one} and subsequently extended in \cite{elliptic_two}. Additionally, some mode-matching proposals can be found in the literature \cite{modematching_elliptical1, modematching_elliptical2, modematching_elliptical3}, normally oriented to analyze discontinuities in an elliptical waveguide. As detailed in Appendix B, the modal functions $\mathbf{\Phi_i}$ involved in the mode-matching are expressed in terms of Mathieu functions \cite{mathieu}. They are solutions of Mathieu's differential equation, the governing wave equation of a homogeneous, lossless elliptical waveguide.  

After obtaining the modal functions, the next step is to calculate the coefficients $\alpha_{ri}$, as expressed in (7). For this purpose, we need to find $I_{ri}$ and calculate the Fourier transforms of the modal functions $\widetilde{\mathbf{\Phi}}_\mathbf{i}=\widetilde{\phi}^x_i \mathbf{\hat{x}}+\widetilde{\phi}^y_i \mathbf{\hat{y}}$ to compute equation \eqref{beta}. 

The two-dimensional Fourier transform of the modal functions is defined as
\begin{equation}
	\widetilde{\mathbf{\Phi}}_\mathbf{i}(k_{x,p},k_{y,s})= \int_{S_{hole}} \mathbf{\Phi_i} \, e^{j(k_{x,p}x+k_{y,s}y)}\, dS.
\end{equation}
The differential surface $dS$ in an elliptic coordinate system can be expressed as $dS=h_{\xi}^2\, d\xi d\eta$, since the scale factor of the $\xi$ and $\eta$ coordinates is the same (see Appendix A). From Appendices A and B, we know how both $\phi^x_i$ and $\phi^y_i$ components of the modal functions can be derived from the $\phi^\xi_i$ and $\phi^\eta_i$ components. Thus, the components of the Fourier transform are written as   
\begin{equation} \label{phitilde_x}
\begin{split}
\widetilde{\phi}_\mathbf{i}^x= h \int_{\xi=0}^{\xi_0} \int_{\eta=0}^{2\pi}  (\phi_i^{\xi}  \sinh &\xi \cos \eta  -  \phi_i^{\eta} \cosh \xi \sin \eta )\\ &\times e^{j(k_{x,p}x+k_{y,s}y)}\, h_{\xi}\, d\xi d\eta
\end{split}
\end{equation}
\begin{equation} \label{phitilde_y}
\begin{split}
\widetilde{\phi}_\mathbf{i}^y= h \int_{\xi=0}^{\xi_0} \int_{\eta=0}^{2\pi}  (\phi_i^{\xi} \cosh& \xi \sin \eta  + \phi_i^{\eta} \sinh \xi \cos \eta)\\ &\times e^{j(k_{x,p}x+k_{y,s}y)}\, h_{\xi}\, d\xi d\eta.
\end{split}
\end{equation}
where $h=\sqrt{a^2-b^2}$. In addition, the inner product of the $r$-th and $i$-th modal functions is expressed as
\begin{equation} \label{inner_product}
	I_{ri}= \int_{S_\mathrm{hole}} \mathbf{\Phi_r} \cdot \mathbf{\Phi_i}\, dS= \int_0^{\xi_0} \int_0^{2 \pi} \left(\phi_{r}^{\xi} \phi_{i}^{\xi}  +\phi_{r}^{\eta} \phi_{i}^{\eta} \right) h_{\xi}^2 d\xi d\eta.
\end{equation}
For the case of perfect metallic waveguides, the modal functions are orthogonal regardless of the geometry of the hole \cite{pozar}. Thus, the inner product $I_{ri}$ is zero when $r\neq i$ and non-zero when $r= i$. This implies that $I_{ri}$ is a diagonal matrix.

The integrals \eqref{phitilde_x}--\eqref{inner_product} are computed with Gaussian quadrature sums \cite{gaussian}  with thirty terms. We have considered fifteen modes inside the hole and nine Floquet modes in the gap region. These values guarantee the convergence of the mode-matching method, since the ratio $15/9=1.67$ is in concordance with the estimation discussed in \cite{fateme2h}. However, more Floquet modes were used to accurately compute the results for small values of the ratio $2a/d$ \cite{guido}.

 As a difference with respect to \cite{guido} and \cite{fateme2h}, this code has been partially optimized when finding the zeros of the determinant of the coefficient matrix in \eqref{linear_system} by taking advantage of the physical properties of the modes. As seen in Figs. \ref{complete}--\ref{eccentricity}, the dispersion diagrams are always continuous functions without an abrupt change. Additionally, the dispersion diagram of the first mode is always below the light line. Thus, by simply searching the frequency range between the frequency associated with the previous point and the one associated with the light line at a particular wavenumber, the computation time of the code is significantly reduced.

\section{Results}
In this section, we study the anisotropic behavior of the unit cell  for different values of the parameters of the elliptical hole. In order to check the validity of the results, all dispersion analyses are compared with the commercial software \textit{CST Microwave Studio}. We will use a reference unit cell to study the individual effect of each parameter in the equivalent refractive index. The dimensions of the reference unit cell are:

\begin{figure}[t]
	\vspace*{-0.2cm}
	\hspace*{-0.3cm}
	\includegraphics[width=0.54 \textwidth]{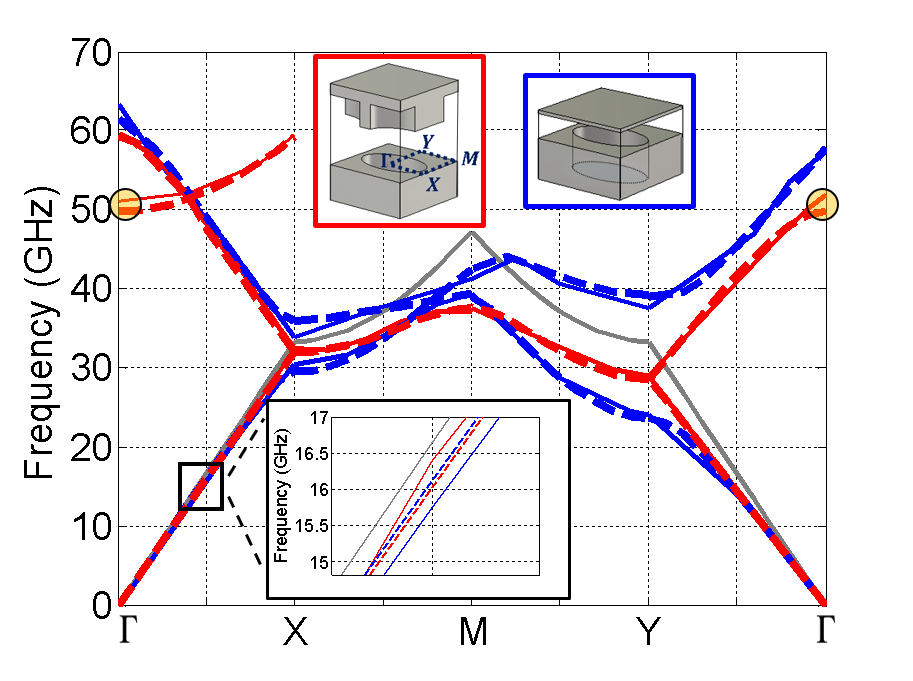}
	\centering
	\caption{\small Dispersion diagram of the irreducible Brilloin zone for the glide-symmetric (\textit{red lines}) and non-glide-symmetric (\textit{blue lines}) unit cells, obtained with the proposed mode-matching (\textit{solid lines}) and \textit{CST} (\textit{dashed lines}). The gray line represents the light line.}
	\label{complete}
\end{figure}
\begin{figure}[t]
	\vspace*{-0cm}
	\hspace*{-0.2cm}
	\includegraphics[width=0.53 \textwidth]{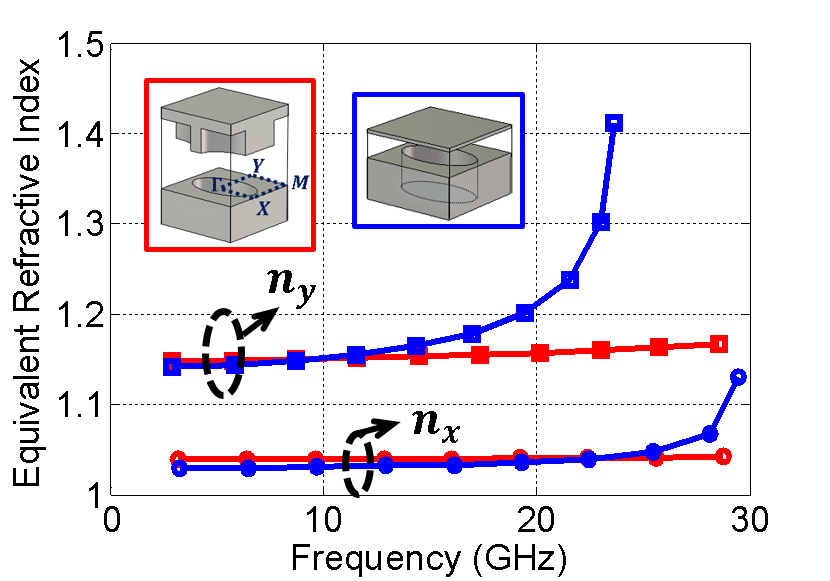}
	\centering
	\caption{\small Equivalent refractive index for wave propagation in $x$- and $y$-directions in the reference glide-symmetric (\textit{red lines}) and non-glide-symmetric unit cells (\textit{blue lines}).}
	\label{nxny_complete2}
\end{figure}

\begin{figure}[!h]
	\vspace*{-0cm}
	\centering
	\subfloat[\hspace*{-0cm}]{
		\hspace*{-0.4cm}
		\label{mu_real}
		\includegraphics[width=0.262\textwidth]{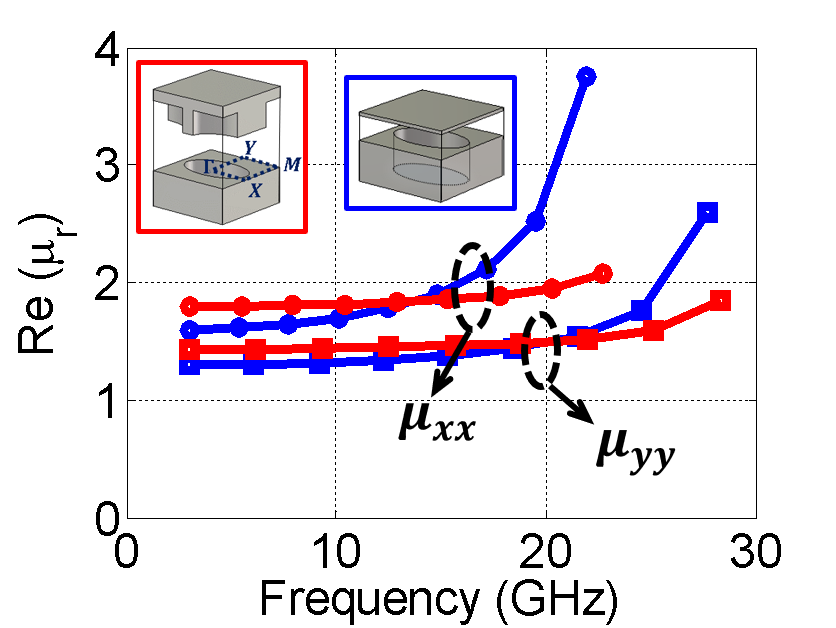}}
	\subfloat[\hspace*{-0.5cm}]{
		\hspace*{-0.6cm}
		\label{eps_real}
		\includegraphics[width=0.26\textwidth]{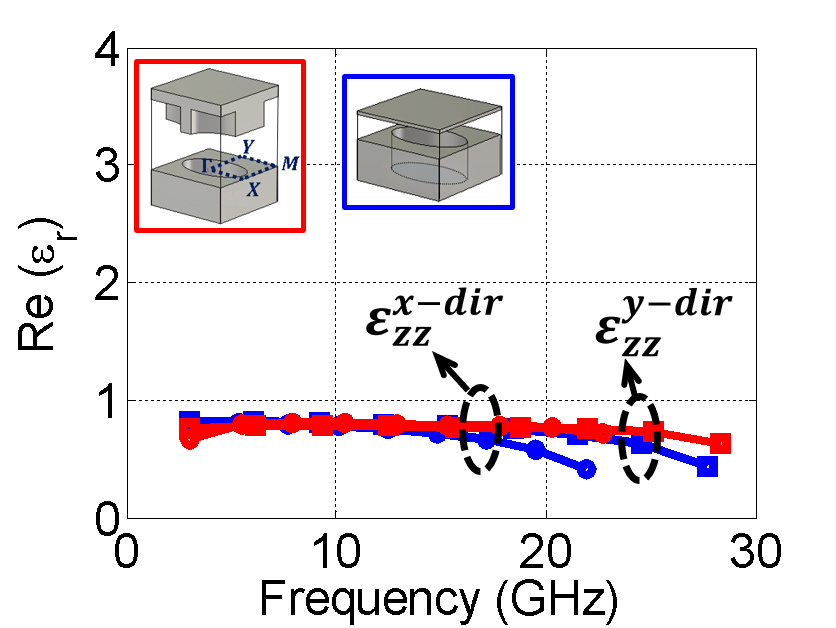}}\\
	\caption*{\small Fig. 4: Constitutive parameters in the reference glide-symmetric (\textit{red lines}) and non-glide-symmetric (\textit{blue lines}) unit cells : (a) Real  part of the permeability. (b) Real  part of the permittivity. }
	\label{epsilon_mu}
\end{figure}

\hspace*{-0.36cm}period $d=4.5$ mm, eccentricity $e=0.8$, gap $g=0.5$ mm, semi-major axis $a=1.9$ mm, and hole depth $h_{\mathrm{hole}}=1.5$ mm. In addition, the gap region and the holes are filled with air ($\varepsilon_r = \varepsilon_{r\mathrm{hole}} = 1$).

\setcounter{figure}{4}    
 Fig. \ref{complete} shows the dispersion diagram over the irreducible Brillouin zone for the glide-symmetric (red) and non-glide-symmetric (blue) reference unit cells. There  an excellent  agreement between the proposed mode-matching (solid lines) and the commercial software \textit{CST} (dashed lines). These results  demonstrate the advantages of using glide symmetry. Due to the vanishing of the stopband between the first and the second modes, the frequency dispersion of the first mode is strongly reduced. The higher linearity of the red curves in the $\Gamma$-X and Y-$\Gamma$ intervals, compared to the blue curves, demonstrates that the equivalent refractive index is almost invariant  over a wider bandwidth in the case of the glide-symmetric configuration (Fig. \ref{nxny_complete2}). This fact is highly desired for the design of wideband lenses. The third mode of the glide-symmetric configuration has also been plotted in Fig. \ref{complete}. This mode is the continuation of the second mode at the ending point of the Y-$\Gamma$ interval (50 GHz), as the yellow circle indicates. Due to the anisotropic behavior of the unit cell, the third mode crosses under the second mode in the $\Gamma$-X interval. For the sake of conciseness, only the first and second modes will be plotted in the next figures. 

Fig. \ref{nxny_complete2} represents  the variation of the equivalent refractive index of both glide-symmetric (red curves) and non-glide-symmetric (blue curves) structures. As previously discussed, the operational bandwidth of the non-glide-symmetric case is narrower compared with the glide-symmetric case. Furthermore, the anisotropy of this structure can be visualized in the figure ($n_y \neq n_x$). The wave that propagates along the $x$-direction ($k_{y,0}=0$) is less disturbed by the presence of the hole. Thus, the equivalent refractive index is smaller in the $x$-direction than in the $y$-direction. Although it is not plotted in Fig. \ref{nxny_complete2}, our code is able to calculate the equivalent refractive index in other propagation directions different from $x$ and $y$. For instance, the equivalent refractive index associated to the main diagonal (45$^{\circ}$) of the Brillouin zone is calculated by setting $0\leq k_{x.0}=k_{y,0}\leq\pi/d$.

Fig. 4 illustrates the constitutive parameters of the glide-symmetric (red) and non-glide-symmetric (blue) unit cells. Following \cite{constitutive_parameters},  the scattering parameters can be used to extract the constitutive parameters of the structures under study. A single row of unit cells (periodic along $y$ direction) is utilized for this purpose. A plane wave that propagates in the parallel-plate along $x$ direction and impinges the row of unit cells is used to excite the structure and extract the S-parameters. For 2-D periodic structures, the permeability components along $x$ and $y$ directions ($\mu_{xx}$ and $\mu_{yy}$) must be extracted separately. Depending on the orientation of the elliptical holes, we derive from the S-parameters the sets \{$\mu_{xx}(f), \varepsilon_z(f)$\} and \{$\mu_{yy}(f), \varepsilon_z(f)$\}. The first set is obtained if the semi-major axis of all elliptical holes that conform the mentioned row is oriented along $y$ direction. The latter is obtained if the semi-major axis of the elliptical holes  is oriented along $x$ direction. Note that $\varepsilon_z$ is obtained at the same time from both configurations, since the electric field is oriented along $z$ direction in the parallel-plate waveguide. When simulating the structures in \textit{CST}, two parallel-plate sections are added at both sides of the row (in $x$ direction) in order to deembed the ports and extract the S-parameters that lead to the constitutive parameters.

The constitutive parameters represent the transverse components of the permeability and the normal component of the permittivity for a wave propagation in $x$ and $y$-directions. It is important to remark that the refractive indexes are cross-related with the constitutive parameters ($ n_x \propto \sqrt{\mu_{yy}}$, $n_y \propto \sqrt{\mu_{xx}}$), as later depicted in Section IV. As observed in Fig. 4, the permeability is higher than the permittivity, and the latter remains invariant for both propagating directions. This fact indicates that the unit cell has a strong magnetic behavior. Therefore, the anisotropy and the variations in the equivalent refractive index (Fig. \ref{nxny_complete2}) are mainly due to the variations in the effective permeability.

\begin{figure}[t]
	\vspace*{-0.15cm}
	\hspace*{-0.2cm}
	\includegraphics[width=0.52 \textwidth]{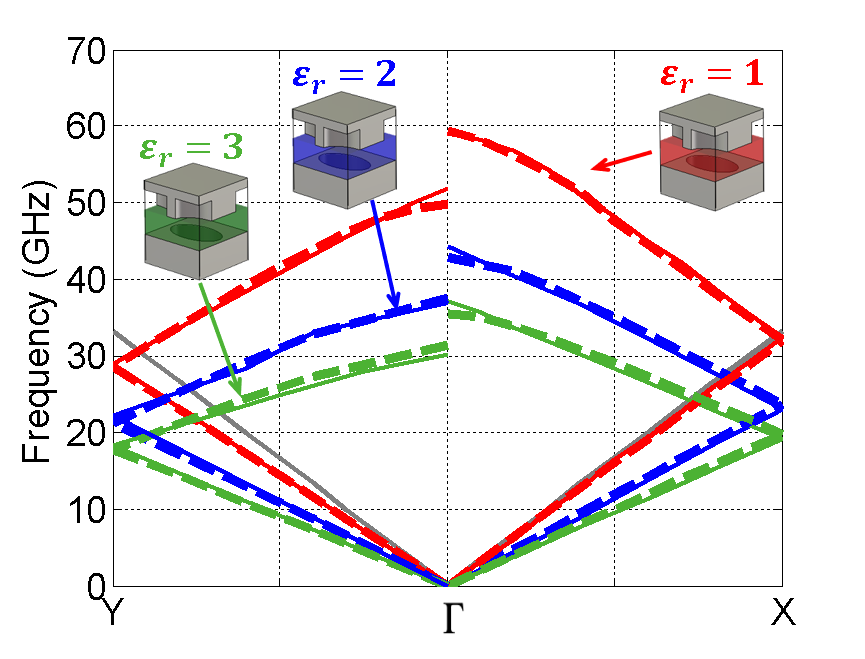}
	\centering
	\caption{\small Dispersion diagram of the glide-symmetric unit cell for different values of the relative permittivity in the gap region, obtained with the proposed mode-matching (\textit{solid lines}) and \textit{CST} (\textit{dashed lines}). The other geometrical parameters of the unit cell are $d=4.5$ mm, $e=0.8$, $g=0.5$  mm, $a=1.9$ mm, and $h_{\mathrm{hole}}=1.5$ mm. The gray line represents the light line. }
	\label{permittivity_png}
\end{figure}

\begin{figure}[t]
	\vspace*{-0.4cm}
	\hspace*{-0.2cm}
	\includegraphics[width=0.52 \textwidth]{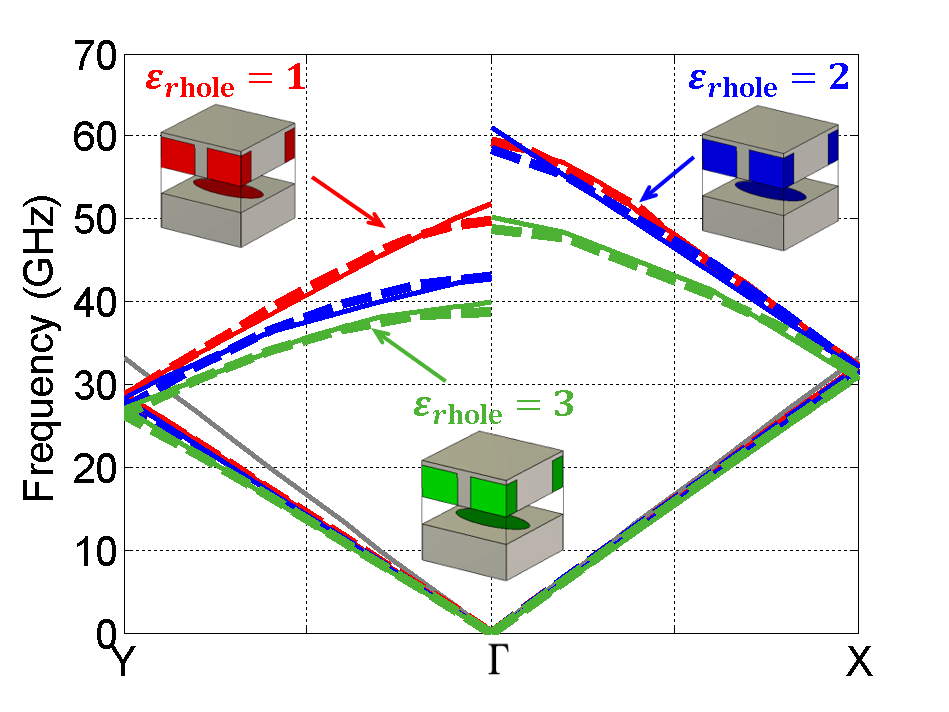}
	\centering
	\caption*{\small Fig. 6: Dispersion diagram of the glide-symmetric unit cell for different dielectrics inside the hole, obtained with the proposed mode-matching (\textit{solid lines}) and \textit{CST} (\textit{dashed lines}). The other geometrical parameters of the unit cell are $d=4.5$ mm, $e=0.8$, $g=0.5$  mm, $a=1.9$ mm, and $h_{\mathrm{hole}}=1.5$ mm. The gray line represents the light line. }
	\label{dielectric_hole}
\end{figure}

In Fig. \ref{permittivity_png}, the variation of the propagation constant as a consequence of filling the gap region with a material of relative permittivity $\varepsilon_r$ is studied over the $\Gamma$-X (propagation in the $x$-direction, $k_{y,0}=0$) and $\Gamma$-Y (propagation in the $y$-direction, $k_{x,0}=0$) intervals. Since it is only necessary to represent the Y-$\Gamma$-X interval to qualitatively observe the presence of the anisotropy, for the sake of conciseness, the interval X-M-Y has not been plotted. As expected, the higher the permittivity, the denser the structure is, so the modes move away from the light line. Beyond that, the frequency dispersion does not seem to increase as the permittivity increases.

\setcounter{figure}{6}    

\begin{figure}[t]
	\vspace*{-0.05cm}
	\hspace*{-0.2cm}
	\includegraphics[width=0.515 \textwidth]{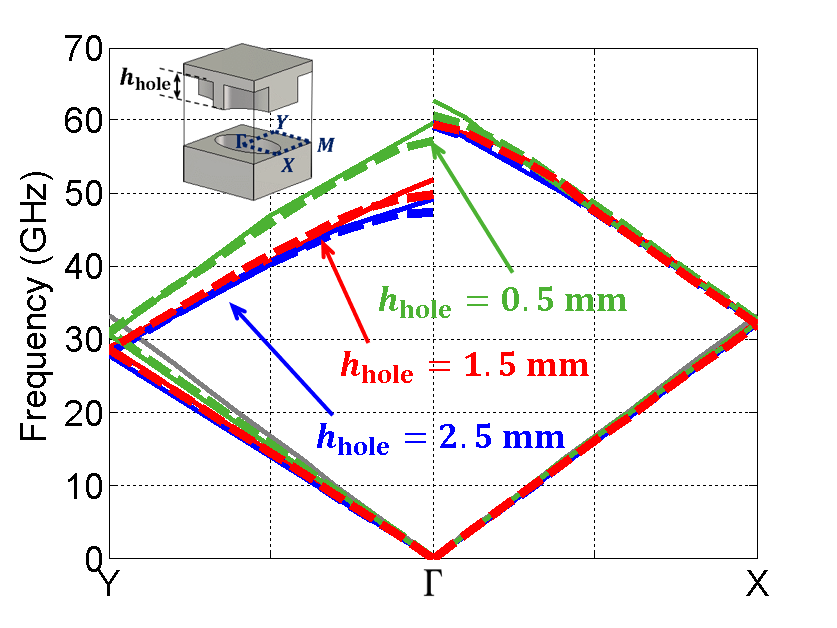}
	\centering
	\caption{\small Dispersion diagram of the glide-symmetric unit cell for different values of the hole depth, obtained with the proposed mode-matching (\textit{solid lines}) and \textit{CST} (\textit{dashed lines}). The other geometrical parameters of the unit cell are $d=4.5$ mm, $e=0.8$, $g=0.5$ mm, $a=1.9$ mm. The gray line represents the light line.}
	\label{hole_png}
\end{figure}

Fig. 6 presents the influence of filling the elliptical hole with a dielectric of relative permittivity $\varepsilon_{r\mathrm{hole}}$ on the propagation constant. Similar to Fig. \ref{permittivity_png}, the equivalent refractive index increases as the hole is filled with denser materials. Additionally, the structure is also more frequency dispersive above 30 GHz as the relative permittivity of the dielectric is higher.

In Fig. \ref{hole_png}, the variation of the hole depth in the glide-symmetric unit cell is studied. For the three different cases considered  here ($h_{\mathrm{hole}}=0.5$ mm, $h_{\mathrm{hole}}=1.5$ mm and $h_{\mathrm{hole}}=2.5$ mm), the agreement between our formulation and \textit{CST} is noticeable. Two phenomena are highlighted in the figure. The results demonstrate that the equivalent refractive index increases as the hole depth increases too. This fact has been recently exploited in the work of \cite{lens} to design a fully-metallic lens for 5G applications. However, there is a limit where the hole depth has no longer influence over the dispersion diagram (blue curve), as also pointed out in \cite{fateme2h}, since the modes exponentially attenuate inside the hole. Furthermore, the unit cell  presents an anisotropic behaviour. Note that by changing the hole depth, the equivalent refractive index has a small variation in the $\Gamma$-X interval. However, the equivalent refractive index varies remarkably in the Y-$\Gamma$ interval. This anisotropy can be used to produce the optically transformed lens presented in Section IV.

\begin{figure}[t]
	\hspace*{-0.35cm}
	\includegraphics[width=0.51 \textwidth]{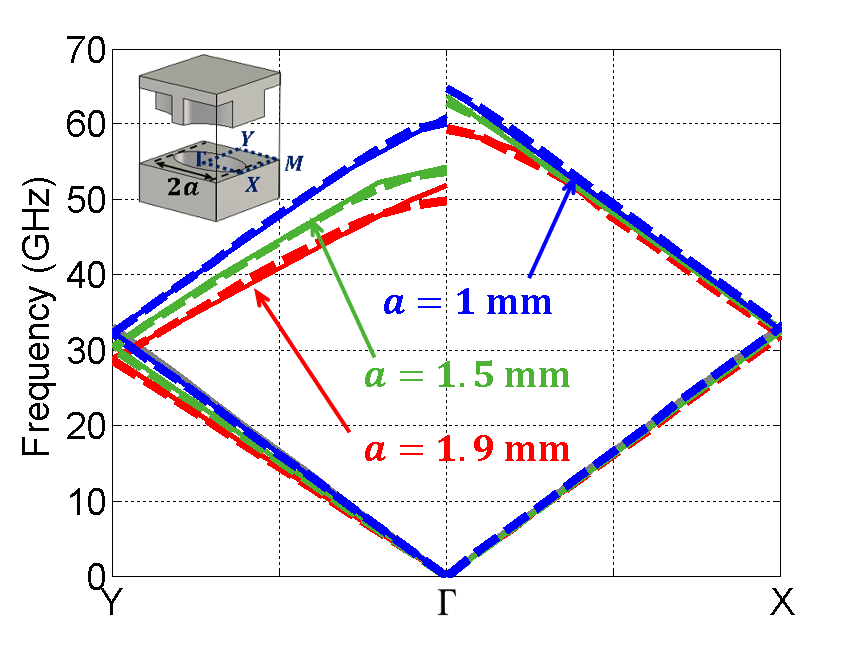}
	\centering
	\caption{\small Dispersion diagram of the glide-symmetric unit cell for different values of the semi-major axis of the hole, obtained with the proposed mode-matching (\textit{solid lines}) and \textit{CST} (\textit{dashed lines}). The other geometrical parameters of the unit cell are $d=4.5$ mm, $e=0.8$, $g=0.5$ mm, and $h_{\mathrm{hole}}=1.5$ mm. The gray line represents the light line.}
	\label{semimajor}
\end{figure}
\begin{figure}[t]
	\vspace*{-0.3cm}
	\hspace*{-0.2cm}
	\includegraphics[width=0.52 \textwidth]{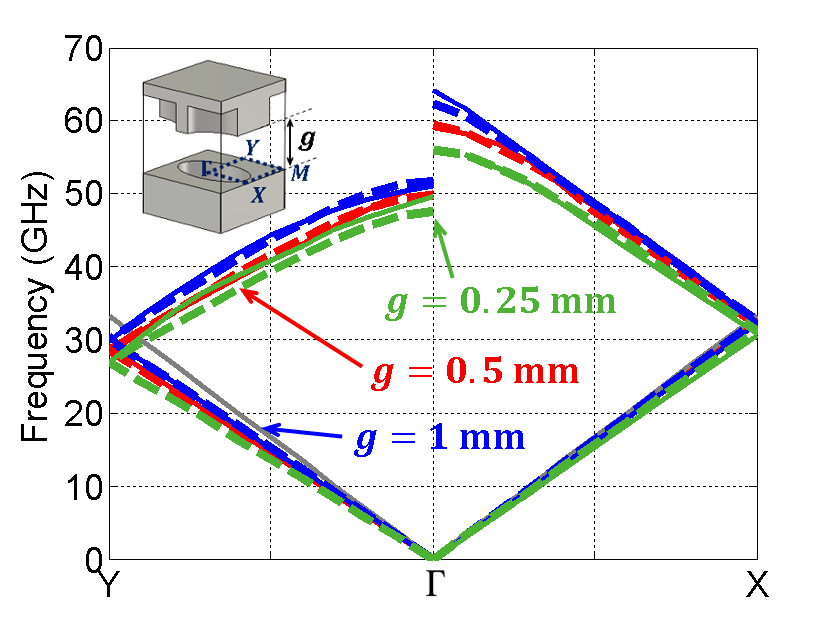}
	\centering
	\caption{\small Dispersion diagram of the glide-symmetric unit cell for different values of the gap, obtained with the proposed mode-matching (\textit{solid lines}) and \textit{CST} (\textit{dashed lines}). The other geometrical parameters of the unit cell are $d=4.5$ mm, $e=0.8$, $a=1.9$ mm, and $h_{\mathrm{hole}}=1.5$ mm. The gray line represents the light line.}
	\label{gap}
\end{figure}

Fig. \ref{semimajor} represents  the variation of the propagation constant  for different values of the semi-major axis $a$ of the ellipse. Three cases are plotted: $a=1.9$ mm, $a=1.5$ mm and $a=1$ mm. The equivalent refractive index increases in the Y-$\Gamma$ interval as the semi-major axis of the ellipse increases too. This is also related with the cross section of the hole: the bigger the hole is, the greater the interaction of the propagating wave with the hole. Thus, the equivalent refractive index approaches the unity as the hole shrinks and the structure becomes a common parallel-plate waveguide.

\begin{figure}[t]
	\vspace*{-0.1cm}
	\hspace*{-0.2cm}
	\includegraphics[width=0.52 \textwidth]{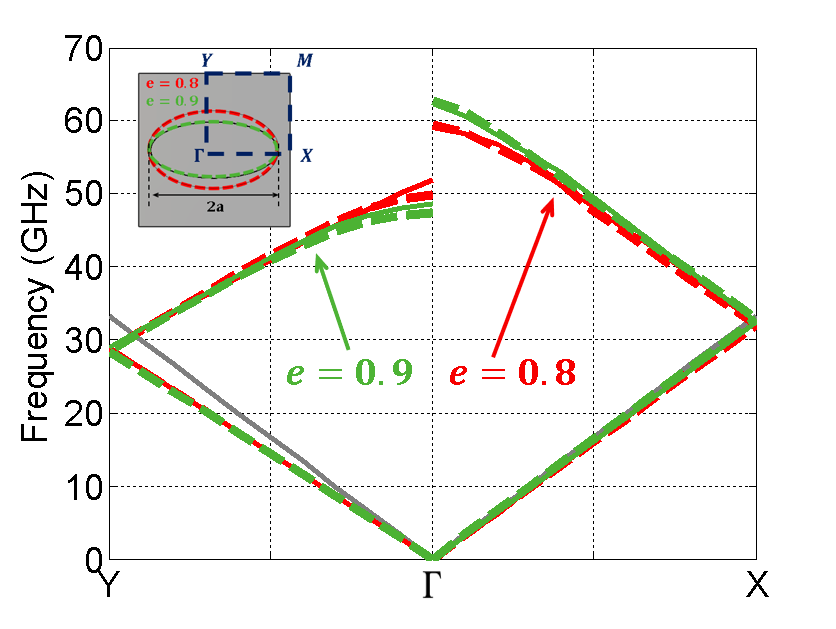}
	\centering
	\caption{\small Dispersion diagram of the glide-symmetric unit cell for different eccentricities of the hole, obtained with the proposed mode-matching (\textit{solid lines}) and \textit{CST} (\textit{dashed lines}). The other geometrical parameters of the unit cell are $d=4.5$ mm, $g=0.5$ mm, $a=1.9$ mm, and $h_{\mathrm{hole}}=1.5$ mm. The gray line represents the light line.  }
	\label{eccentricity}
\end{figure}

\begin{figure}[t]
	\vspace*{-0.3cm}
	\hspace*{-0.35cm}
	\includegraphics[width=0.54 \textwidth]{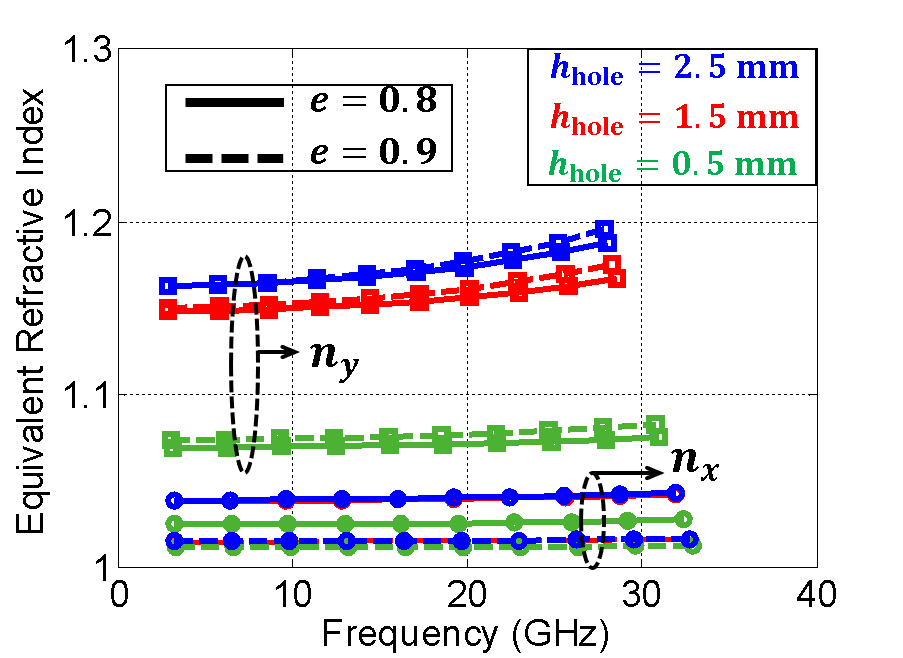}
	\centering
	\caption{\small Equivalent refractive index of the glide-symmetric unit cell for wave propagation in $x$- and $y$-directions. The geometrical parameters of the unit cell are $d=4.5$ mm, $g=0.5$ mm, and $a=1.9$ mm.}
	\label{anisotropy}
	\end{figure}

In Fig. \ref{gap}, the influence of the gap height on the propagation constant is studied. In this case, three different values are considered: $g=1$ mm, $g=0.5$ mm and $g=0.25$ mm. Even though the gap height is not a parameter to be typically tuned for lens designs, its influence on the equivalent refractive index cannot be neglected. Some discrepancies between our formulation and \textit{CST} appear for really small values of $g$ such as $g=0.25$ mm, as more than fifteen modes need to be considered here for accurate results. As expected, the equivalent refractive index becomes higher as the gap height is smaller, since the interaction between the top and bottom metallic layers is stronger.

The variation of the propagation constant as a consequence of modifying the eccentricity of the hole is  studied in Fig. \ref{eccentricity}. The design of anisotropic lenses may require very flattened ellipses, so the effect of the anisotropy is highly marked in the structure. Thus, two high-eccentricities values are considered here: $e=0.8$ and $e=0.9$. As the eccentricity of the hole increases, the modes in the dispersion diagram approach the light line.


Finally, Fig. \ref{anisotropy} presents a parametric study of the variation in the equivalent refractive index of the glide-symmetric unit cell when modifying the hole depth and the eccentricity of the ellipse. The potential of  glide-symmetric structures with elliptical holes is reflected in the figure. Firstly, flattened ellipses easily control the anisotropy of the structure in an easy manner, since $n_x$ is near to one and $n_y$ varies. Secondly, the reduction of the  frequency dispersion in the first mode is appreciable in Fig. \ref{anisotropy}. 

\section{Compressed Maxwell Fish-Eye Lens}
In this section, we take advantage of both the broadband response due to glide symmetry and the anisotropy related to elliptical holes to produce a broadband compressed Maxwell fish-eye lens (MFE). Thus, we present a proof-of-concept design with a compression factor of 33.33\% and wideband capabilities.

The Maxwell fish-eye lens is a rotationally-symmetric graded-index lens that focuses any point-source excitation to the opposite side of the circle \cite{maxwell, maxwell_rhiannon}. The required refractive index for this lens is given by:
\begin{equation}
	n(r)=\frac{2n_0}{1+(r/R)^2} 
\end{equation}
where $r=\sqrt{x^2+y^2}$, $R$ is the radius of the lens and $n_0$ is the background refractive index, assumed to be one here.

Here, we attempt to compress the lens by a factor $\beta$ while maintaining the same aperture in the other direction. Transformation optics (TO) and the anisotropy related to elliptical holes can be used for our purpose. Essentially, TO relates the electromagnetic fields and constitutive parameters of the original space, namely virtual space ($x,y,z$ coordinates), to the desired transformed space, known as physical space ($x', y', z'$ coordinates). Since we are compressing the lens along the $y$ direction, the coordinates of the transformation are:
\begin{equation}
x'=x, \qquad y'=\frac{y}{\beta}, \qquad z'=z .
\end{equation}
\hspace*{-0.17cm}In the physical space, we are interested in having isotropic materials outside the lens, similar to the virtual space (air: $\varepsilon_0$, $\mu_0$). Thus, the formulation of \cite{mahsa} particularized to the MFE lens leads us to the constitutive parameters:
\begin{subequations} \label{constitutive_parameters}
	\begin{equation}
		\overline{\overline{\varepsilon'}} = \varepsilon_0 \left( {\begin{array}{ccc}
			1 & 0 & 0 \\
			0 & 1/\beta^2 & 0 \\
			0 & 0 & 1 \\
			\end{array} } \right)
	\end{equation}
	\begin{equation}
	\overline{\overline{\mu'}} = \mu_0\, n^2(r') \left( {\begin{array}{ccc}
		1 & 0 & 0 \\
		0 & 1/\beta^2 & 0 \\
		0 & 0 & 1 \\
		\end{array} } \right) 
	\end{equation}	
\end{subequations}
where $r' = \sqrt{x'^2 + (\beta y')^2}$. From the material properties of the physical space, the refractive index can be calculated:
\begin{subequations} \label{refractive_index}
	\begin{equation}
	n_x'=\sqrt{\varepsilon_{zz}' \mu_{yy}'}
	\end{equation}
	\begin{equation}
	n_y'=\sqrt{\varepsilon_{zz}' \mu_{xx}'}\, .
	\end{equation}
\end{subequations}
Note that, since the refractive index is cross-related with the constitutive parameters ($ n_x' \propto \sqrt{\mu_{yy}'}$, $n_y' \propto \sqrt{\mu_{xx}'}$) and $\beta>1$, it follows from equations \eqref{constitutive_parameters} and \eqref{refractive_index} that $n_x'<n_y'$. This is, the refractive index is higher along the compressed direction. Therefore, the elliptical holes must be oriented as shown in Fig. 12.

\begin{figure}[t]
	\centering
	\subfloat[\hspace*{-0cm}]{
		\label{maxwell_rogers}
		\includegraphics[width=0.45\textwidth]{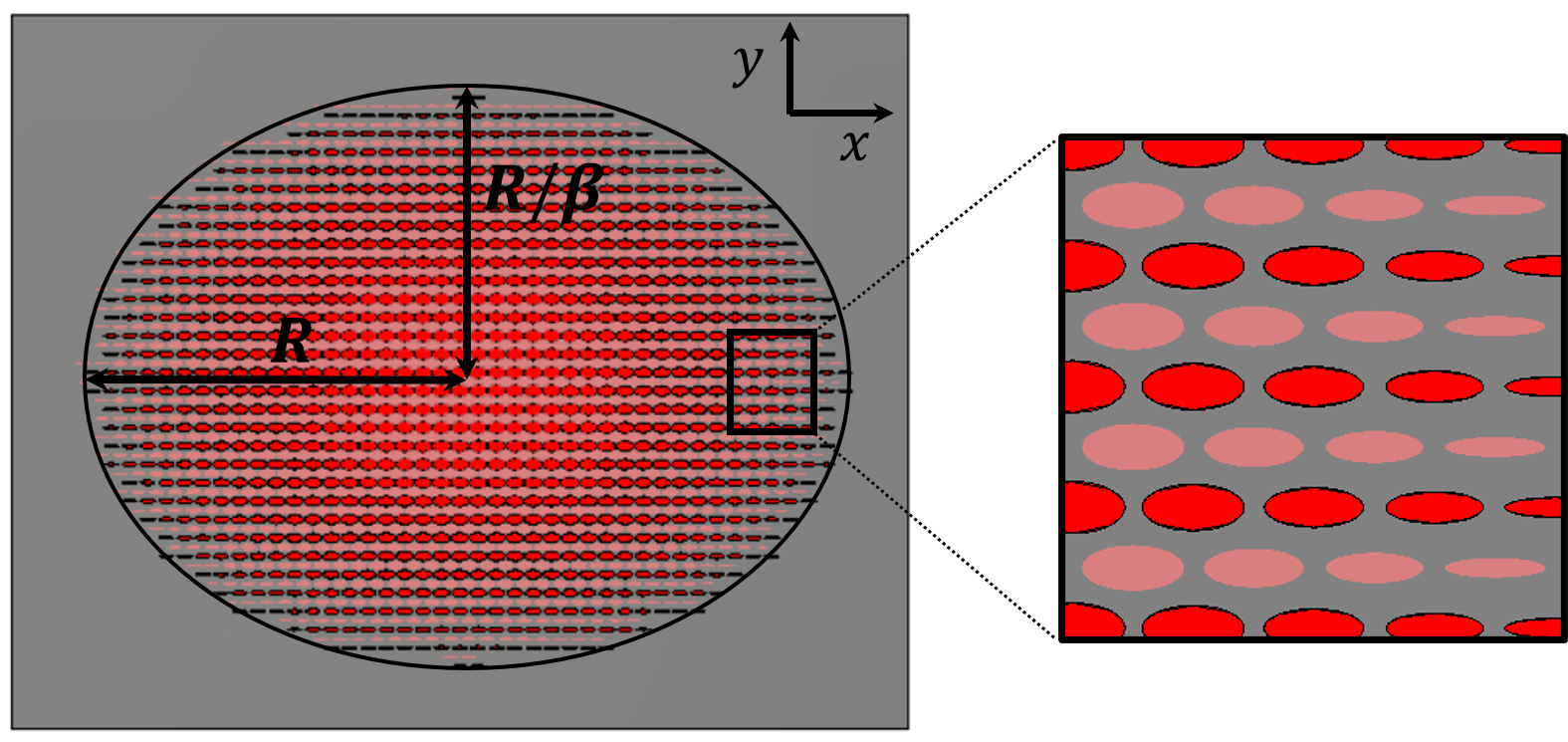}}\\
	\vspace*{-0.2cm}
	\subfloat[\hspace*{-0.6cm}]{
		\hspace*{-0.6cm}
		\label{unitcell_glide_rogers}
		\includegraphics[width=0.24\textwidth]{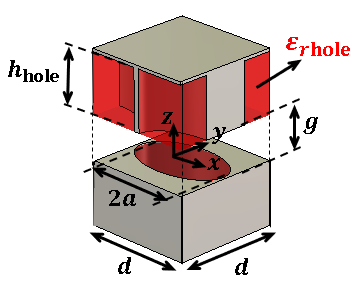}}
	\caption*{\small \textcolor{black}{Fig. 12: (a) Top view of the compressed Maxwell fish-eye lens with a metasurface detail  and (b) glide-symmetric unit cell. Elliptical holes are filled with a dielectric of relative permittivity $\varepsilon_{r\mathrm{hole}}$.}}
	\label{maxwell}
\end{figure}

The schematic for the compressed MFE lens is shown in Fig. 12. Bright and mat red colors relate the holes of the bottom and top layers, respectively. In order to compare our results with \cite{maxwell_oscar}, the compression factor is chosen to be $\beta = 4/3$ (33.33 \%). The radius of the lens is set to $R=9$ cm. As illustrated in Fig. 12(b), we have placed a dielectric of relative permittivity $\varepsilon_{r\mathrm{hole}}$ inside the elliptical holes to increase the equivalent refractive index in both $x$ and $y$ directions and reach the required values. 


The map of the required refractive indexes is displayed in Fig. 13. Some of the values extracted from equation \eqref{refractive_index} are less than one. In particular, the minimum index associated to the propagation in the $x$-direction is $n_{x,\mathrm{min}}'=1/\beta = 0.75$. Since the proposed unit cell cannot reach these values, the map plot of Fig. 13 was normalized by $1/\beta$ so the minimum required index is one.

Fig. 14 shows the equivalent refractive index obtained at 10 GHz with the unit cell of Fig. 12(b) for different hole sizes when filled with Rogers-3010 dielectric ($\varepsilon_{r\mathrm{hole}}=10.2$). In order to implement the compressed MFE lens, for each spatial point, we must select from Fig. 14 the values of the eccentricity and major-axis that match the equivalent refractive indexes closest to the map of Fig. 13. In the center of the lens, the required refractive indexes are higher. As a consequence, the holes are bigger there. 

Subsequently, the compressed MFE lens is realized and simulated in \textit{CST Microwave Studio}. The normalized absolute value of the electric field is displayed in Fig. 15 from 2.5 GHz to 10 GHz for three excitation points at 0$^\mathrm{o}$, 45$^\mathrm{o}$ and 90$^\mathrm{o}$. As observed, the compressed lens shows an ultrawideband performance, enhanced by the use of glide symmetry. Since the maximum equivalent refractive index calculated in Fig. 14 is $n_{y,\mathrm{max}}'=2.86$, the maximum achievable compression factor in the glide case will be $\beta_G=43.00$ \%.  On the other hand, a further simulation in \textit{CST} reveals that the non-glide-symmetric version of the unit cell of Fig. 12(b) only offers a maximum compression factor of $\beta_{NG}=3.05$ \%.

\begin{figure}[t]
	\vspace*{-0.5cm}
	\centering
	\subfloat[\hspace*{-0.2cm}]{
		\hspace*{-0.38cm}
		\label{nx}
		\includegraphics[width=0.247\textwidth]{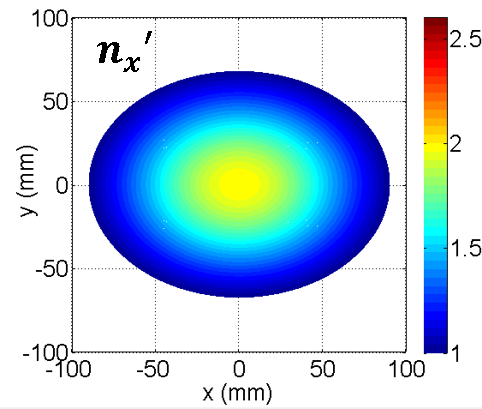}}
	\subfloat[\hspace*{-0.2cm}]{
		\hspace*{-0.2cm}
		\label{ny}
		\includegraphics[width=0.257\textwidth]{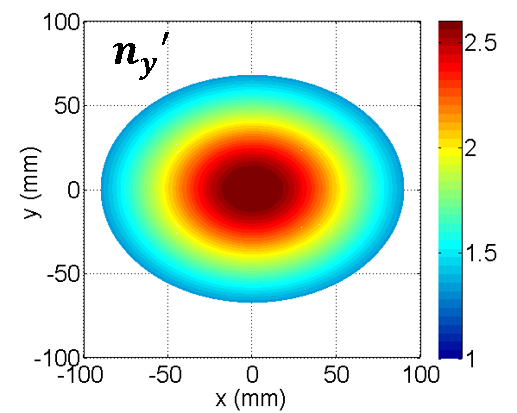}}
	\caption*{\small Fig. 13: Map of refractive indexes after normalization for a 33.33\% compression in the MFE lens: (a) $n_x'$ , (b) $n_y'$.}
	\label{index}
\end{figure}

\begin{figure}[t]
	\vspace*{-0.cm}
	\centering
	\subfloat[\hspace*{-0.2cm}]{
		\hspace*{-0.35cm}
		\label{nx}
		\includegraphics[width=0.254\textwidth]{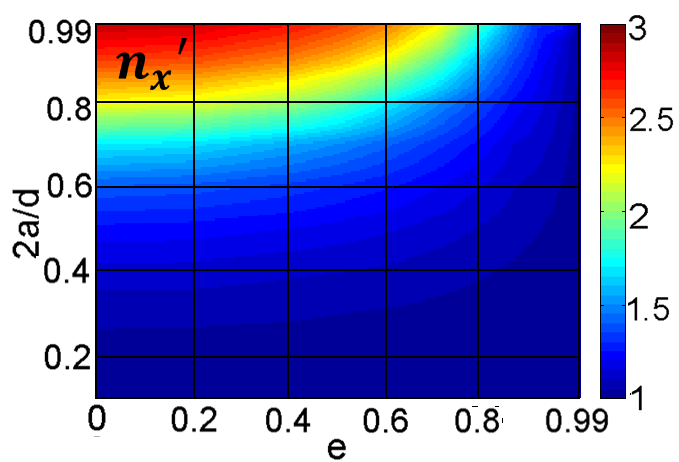}}
	\subfloat[\hspace*{-0.2cm}]{
		\hspace*{-0.2cm}
		\label{ny}
		\includegraphics[width=0.25\textwidth]{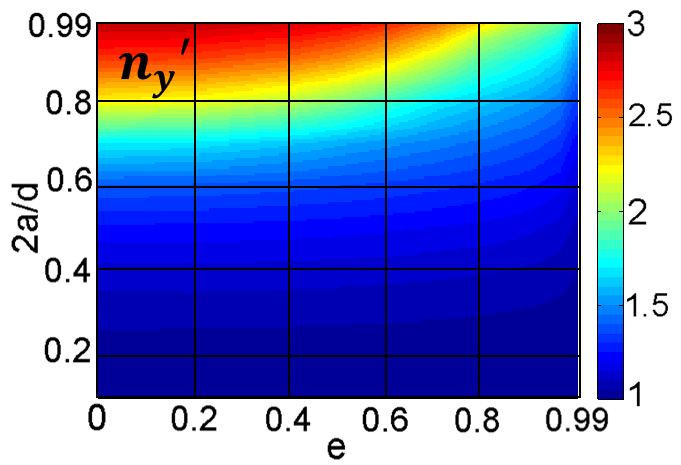}}
	\vspace*{-0cm}
	\caption*{\small Fig. 14: Equivalent refractive index extracted at 10 GHz from the glide-symmetric unit cell when varying the eccentricity and the major-axis of the ellipse: (a) $n_x'$ , (b) $n_y'$. The geometrical parameters of the unit cell are: $d=4.5$ mm, $g=0.1$ mm, and $h_\mathrm{hole}=2.5$ mm.}
	\label{design}
\end{figure}

\begin{figure}[t]
	\centering
	\subfloat[\hspace*{-0.3cm}]{
		\label{maxwell1}
		\includegraphics[width=0.48\textwidth]{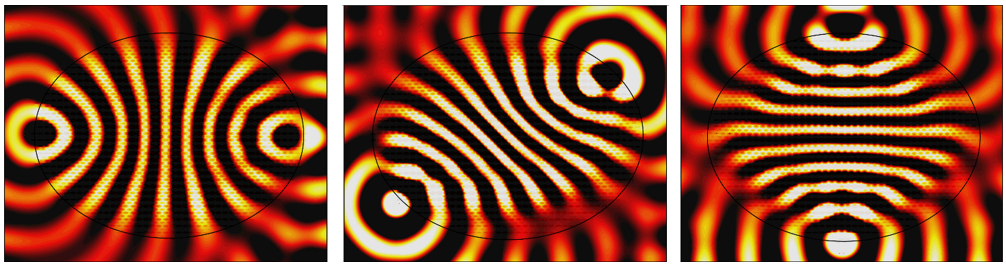}}\\
	\vspace*{-0.3cm}
	\subfloat[\hspace*{-0.3cm}]{
		\label{maxwell2}
		\includegraphics[width=0.48\textwidth]{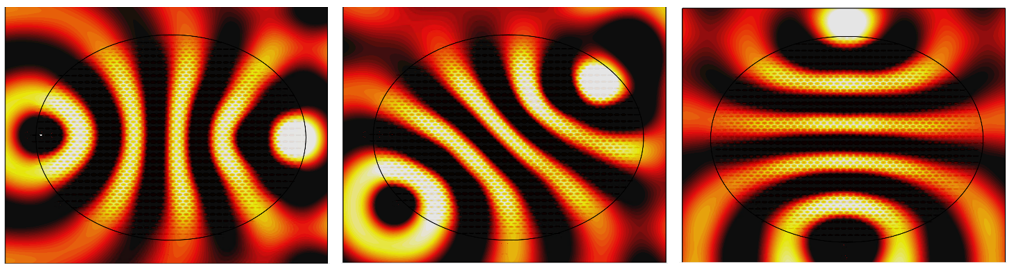}}\\
	\vspace*{-0.3cm}
	\subfloat[\hspace*{-0.3cm}]{
		\label{maxwell3}
		\includegraphics[width=0.48\textwidth]{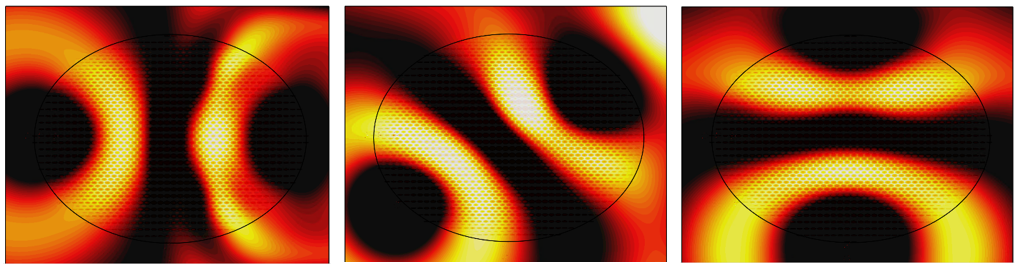}}
	\caption*{\small Fig. 15: Normalized electric field distribution (absolute value) of the compressed MFE lens at different frequencies: (a) 10 GHz, (b) 5 GHz, (c) 2.5 GHz. The point source is located at 0$^\mathrm{o}$, 45$^\mathrm{o}$ and 90$^\mathrm{o}$. The black line indicates the contour of the lens.}
	\label{electric_field}
\end{figure}

\setcounter{figure}{15}

Finally, a performance comparison is made with respect \cite{maxwell_oscar}, where a MFE lens was compressed with a conformal transformation and fully isotropic dielectric materials. The applied compression factor in \cite{maxwell_oscar} is identical to our work, a 33.33 \%. However, their design needs a higher variation of the  refractive index ($2.38/0.68=3.5$) compared to ours ($2.67$) in order to fulfill the same compression requirements. As a conclusion, the use of anisotropic materials such as the elliptical holes reduces the required variation of the refractive index when compared to an isotropic design.


\section{Conclusion}

In this paper, we analyzed the wave propagation in a glide-symmetric metallic structure composed of periodic elliptical holes. For this purpose, a mode-matching based on the generalized Floquet theorem has been applied. The dispersion diagrams of the unit cell have been obtained for the different values of all the geometric parameters. It has been demonstrated that glide-symmetric structures with periodic elliptical holes exhibit an anisotropic refractive index over a wide range of frequencies. Furthermore, it has been confirmed that the refractive index can be tuned in a wider bandwidth in the glide-symmetric case. 


In order to validate the capabilities of the presented structure for practical scenarios, a compressed lens  has been designed. With the use of periodic elliptical holes in a glide-symmetric configuration, the profile of a MFE lens has been compressed of 33.33 \%. The simulated MFE lens is wideband, operating from 2.5 GHz to 10 GHz. It is reported how the use of glide symmetry increases the equivalent refractive index, so higher compression factors can be achieved. For the same dimensions and materials in the unit cell, the non-glide version only offers a maximum compression factor of 3.05 \%, while with glide symmetry, a 43\% is achieved. Finally, a comparison between isotropic and anisotropic materials is done. The use of anisotropic materials substantially reduces the required range of refractive index when compressing the lens. Thus, periodic elliptical holes arranged in a glide-symmetric configuration are a promising candidate for the manufacturing of reduced-size low-loss low-cost wideband planar lenses.


\begin{appendices}
\section{Elliptic Cylindrical Coordinates}

\begin{figure}[t]
	\hspace*{-0.2cm}
	\includegraphics[width=0.4 \textwidth]{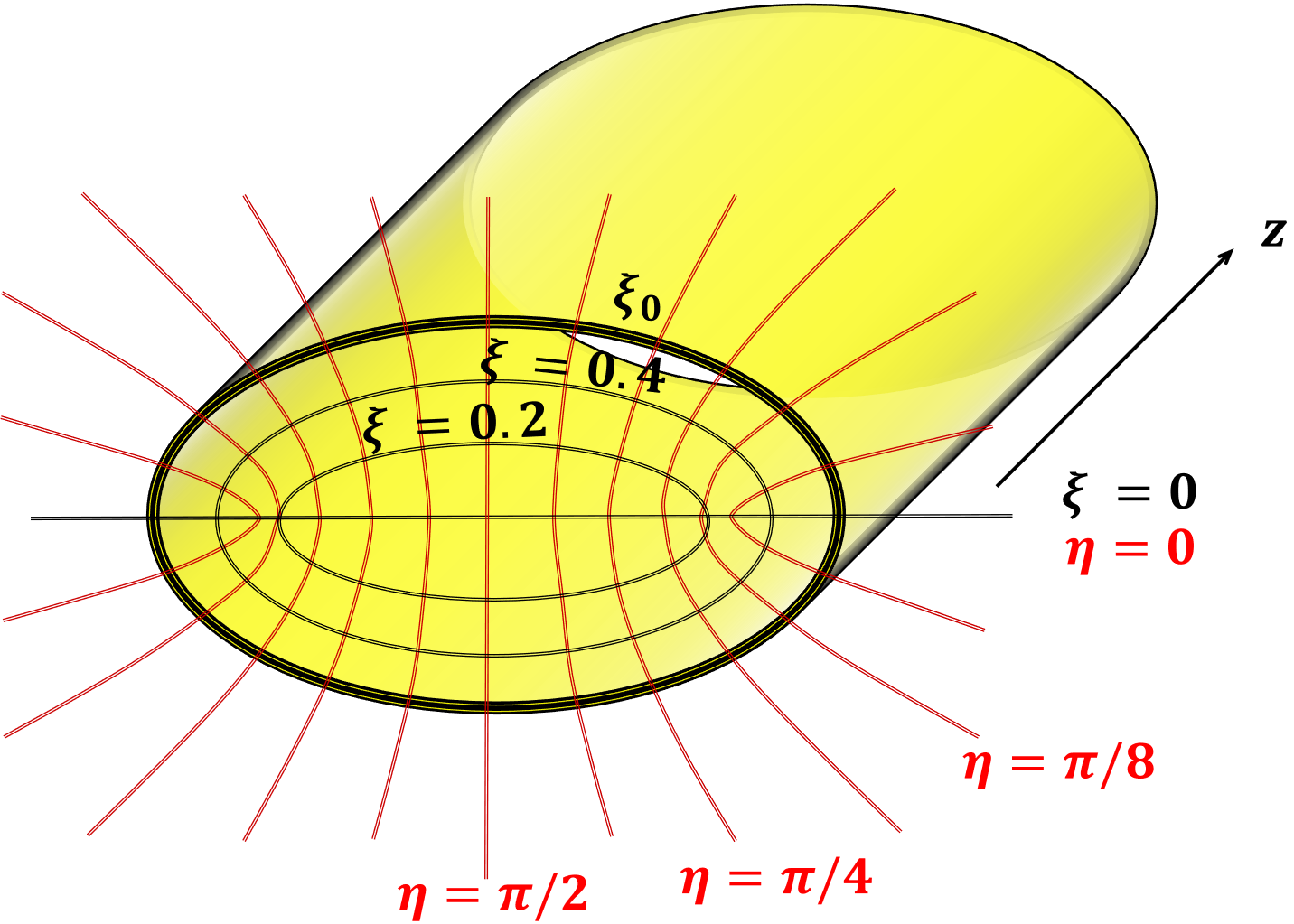}
	\centering
	\caption{\small Elliptic cylindrical coordinate system describing an elliptic waveguide.}
	\label{guia}
\end{figure}

Elliptic cylindrical coordinates, depicted in Fig. \ref{guia}, are defined by a radial coordinate $\xi$, which describes a set of confocal ellipses, an angular coordinate $\eta$, which describes a set of hyperbolae sharing the same foci, and a longitudinal coordinate $z$. Their relation with the cartesian coordinates ($x,y,z$) is \cite{elliptic}:
\begin{equation} \label{cartesian}
\begin{cases}
x=h \cosh  \xi \cos  \eta \\
y=h \sinh  \xi \sin  \eta\\
z=z
\end{cases}
\end{equation}
where $h=ae$ is the focal distance of the ellipse, $a$, $b$ are the semi-major and semi-minor axes of the ellipse, and $e=\sqrt{a^2-b^2}/a$ is the eccentricity of the ellipse. From the definition of a vector $\mathbf{r}=x\mathbf{\hat{x}}+y\mathbf{\hat{y}}+z\mathbf{\hat{z}}$ and the use of \eqref{cartesian}, the basis unit vectors $\boldsymbol{\hat{\xi}}, \boldsymbol{\hat{\eta}}, \mathbf{\hat{z}}$ can be obtained as $\boldsymbol{\hat{\xi}}=\frac{1}{h_\xi} \frac{\partial \mathbf{r}}{\partial \xi}$, $\boldsymbol{\hat{\eta}}=\frac{1}{h_\eta} \frac{\partial \mathbf{r}}{\partial \eta}$, $\mathbf{\hat{z}}=\frac{1}{h_z} \frac{\partial \mathbf{r}}{\partial z}$, where $h_\xi=h_\eta=h\sqrt{\sinh^2 \xi + \sin^2  \eta}$ and $h_z=1$ are the scale factors. Thus, the basis unit vectors in elliptic cylindrical coordinates are expressed as
\begin{gather}
\begin{bmatrix} \boldsymbol{\hat{\xi}}  \\ \boldsymbol{\hat{\eta}} \\ \mathbf{\hat{z}} \end{bmatrix}
= \frac{h}{h_\xi}
\begin{bmatrix}
\sinh \xi \cos \eta    & \cosh \xi \sin \eta & 0\\
-\cosh \xi \sin \eta & \sinh \xi \cos \eta& 0 \\
0& 0 & \frac{h_\xi}{h}
\end{bmatrix}
\begin{bmatrix} \mathbf{\hat{x}}  \\ \mathbf{\hat{y}}  \\ \mathbf{\hat{z}} \end{bmatrix}
\end{gather}

Furthermore, the surface differential $dS$ can be derived by applying the formulas of an orthogonal coordinate system. Thus, 
\begin{equation}
	dS=h_\xi h_\eta d\xi d\eta= h^2 (\sinh^2 \xi + \sin^2  \eta)\, d\xi d\eta.
\end{equation}
Other differential operators as the Laplacian of a scalar field $\nabla^2 \phi$, the gradient of a scalar field $\nabla \phi$, the divergence of a vector field $\nabla \cdot \mathbf{\Phi}$, or the curl of a vector field $\nabla \times \mathbf{\Phi}$ can be derived in the same manner \cite{elliptic}.

\section{Modal Functions in the Elliptical Hole}
As previously discussed, the elliptical hole can be seen as a short-circuited elliptical waveguide. The $i$-th modal function is expressed in terms of elliptical coordinates as $\mathbf{\Phi_i}=\phi_i^\xi \, \boldsymbol{\hat{\xi}} + \phi_i^\eta \, \boldsymbol{\hat{\eta}}$. For TM modes ($Y_i=k_0/(\eta_0\, k_{zi})$):
\begin{equation} \label{modal1}
\phi^{\xi}_{i}(\xi,\eta)= \frac{1}{h_{\xi}}\begin{Bmatrix} Ce_{m}'(\xi,q_{\mathrm{e}mn})\, ce_{m}(\eta,q_{\mathrm{e}mn}) \, \, \, \textrm{(even)}\\   
Se_{m}'(\xi,q_{\mathrm{o}mn})\, se_{m}(\eta,q_{\mathrm{o}mn}) \, \, \, \, \textrm{(odd)}
\end{Bmatrix}
\end{equation}
\begin{equation}
\phi^{\eta}_i(\xi,\eta)= \frac{1}{h_{\xi}}\begin{Bmatrix} Ce_{m}(\xi,q_{\mathrm{e}mn})\, ce_{m}'(\eta,q_{\mathrm{e}mn}) \, \, \, \textrm{(even)}\\
Se_{m}(\xi,q_{\mathrm{o}mn})\, se_{m}'(\eta,q_{\mathrm{o}mn}) \, \, \, \, \textrm{(odd)} 
\end{Bmatrix}
\end{equation}
For TE modes ($Y_i=k_{zi}/(k_0\, \eta_0) $):
\begin{equation}
\phi^{\xi}_i(\xi,\eta)=\frac{1}{h_{\xi}} \begin{Bmatrix}  Ce_{m}(\xi,q'_{\mathrm{e}mn})\, ce'_{m}(\eta,q'_{\mathrm{e}mn}) \, \, \, \textrm{(even)}\\   
Se_{m}(\xi,q'_{\mathrm{o}mn})\, se'_{m}(\eta,q'_{\mathrm{o}mn}) \, \, \, \, \textrm{(odd)} 
\end{Bmatrix}
\end{equation}
\begin{equation} \label{modal4}
\phi^{\eta}_i(\xi,\eta)=\frac{-1}{h_{\xi}} \begin{Bmatrix}  Ce'_{m}(\xi,q'_{\mathrm{e}mn})\, ce_{m}(\eta,q'_{\mathrm{e}mn}) \, \, \, \textrm{(even)}\\
Se'_{m}(\xi,q'_{\mathrm{o}mn})\, se_{m}(\eta,q'_{\mathrm{o}mn}) \, \, \, \, \textrm{(odd)} 
\end{Bmatrix}
\end{equation}
where $ce_{m}(\eta,q_{\mathrm{e}mn})$ and $se_{m}(\eta,q_{\mathrm{o}mn})$ are the even and odd angular Mathieu functions of order $m$, $Ce_{m}(\eta,q'_{\mathrm{e}mn})$ and $Se_{m}(\eta,q'_{\mathrm{o}mn})$ are the even and odd radial Mathieu functions of the first kind of order $m$ \cite{mathieu}. The prime symbol ($'$) denotes the derivative with respect to $\xi$ or $\eta$ in the Mathieu functions. As a difference with the circular waveguide \cite{pozar}, the equations of the tangential fields \eqref{modal1}--\eqref{modal4} show that there are two propagating solutions (even and odd) for each mode. Thus, an additional index is needed to differentiate them. We will use the subscript ``$\mathrm{e}$" to indicate that the $mn$-th mode is even and ``$\mathrm{o}$" to indicate that the $mn$-th mode is odd, being the complete notation TM/TE$_{\mathrm{e}/\mathrm{o}\,mn}$. Note that the even and odd solutions degenerate into a unique solution when the eccentricity of the ellipse is zero and the elliptic waveguide transforms into a circular waveguide.

\begin{figure}[t]
	\includegraphics[width=0.47 \textwidth]{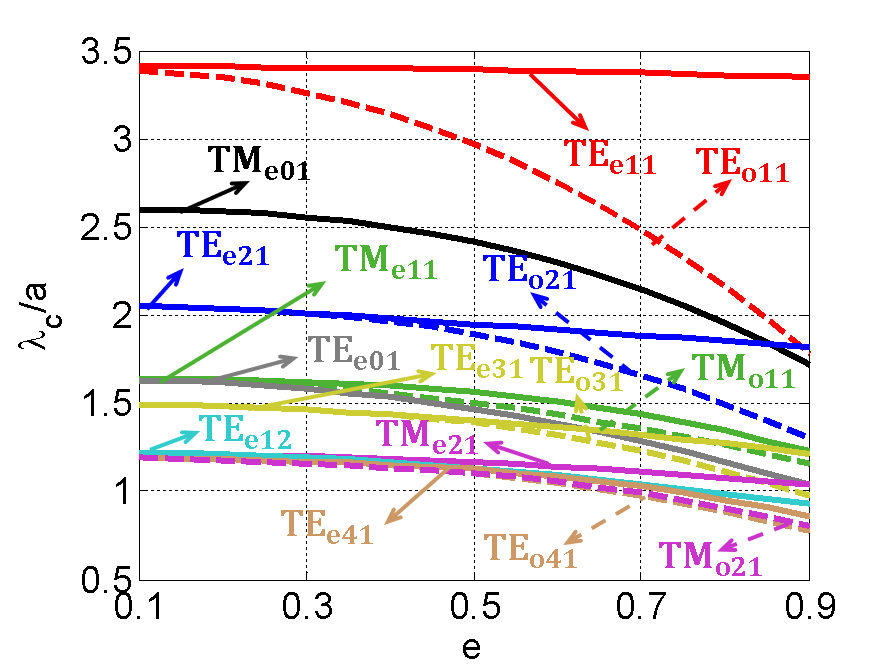}
	\centering
	\caption{\small Normalized cut-off wavelength $\lambda_c/a$ for the first fifteen modes as a function of the eccentricity $e$ of the elliptical hole.  }
	\label{sort_modes}
\end{figure}

The $q$-parameters ($q_{\mathrm{e}mn}$, $q_{\mathrm{o}mn}$, $q'_{\mathrm{e}mn}$ and $q'_{\mathrm{o}mn}$) are calculated by imposing the boundary condition on the metallic wall of the waveguide. They are directly related with the cutoff wavelength $\lambda_c$ in the elliptic waveguide as \cite{elliptic_two}:
\begin{equation} \label{lambdac}
\lambda_c= \frac{\pi a e}{\sqrt{q}}.
\end{equation} 
In the case of TM modes, the parameters $q_{\mathrm{e}mn}$, $q_{\mathrm{o}mn}$ are the $n$-th roots of Mathieu Radial (cosine and sine) functions:
\begin{equation} \label{b1}
\begin{Bmatrix}	Ce_{m}(\xi_0,q_{\mathrm{e}mn}) \\
Se_{m}(\xi_0,q_{\mathrm{o}mn})
\end{Bmatrix}=0 \, .
\end{equation}
In the case of TE modes, the parameters $q'_{\mathrm{e}mn}$, $q'_{\mathrm{o}mn}$ are the $n$-th roots of the first derivative of the Mathieu Radial functions:
\begin{equation} \label{b2}
\begin{Bmatrix}	Ce'_{m}(\xi_0,q'_{\mathrm{e}mn}) \\
Se'_{m}(\xi_0,q'_{\mathrm{o}mn})
\end{Bmatrix}=0 \, .
\end{equation}
The radial value $\xi_0$ delimits the metallic surface of the waveguide. This parameter is related with the eccentricity of the elliptical hole according to the expression $\cosh \xi_0=1/e$ \cite{elliptic_two}. Extracting the $q$-roots from the boundary conditions \eqref{b1}--\eqref{b2}  is not easy, since the Mathieu radial functions $Ce_{m}$ and $Se_{m}$ are expressed as an infinite sum of hyperbolic functions \cite{table_mathieu,vacuum} and dependent on the eccentricity of the hole. The normalized cut-off frequencies of the first fifteen modes inside the elliptical hole, obtained for different values of the eccentricity, are represented in Fig. \ref{sort_modes}. For the sake of clarity, the even and odd versions of the same mode have been plotted with the same color. Additionally, the even modes are displayed with solid lines and the odd modes with dashed lines.

The fundamental mode inside the elliptic hole is the TE$_{\mathrm{e}11}$. However, the order of the higher-order modes depends on the eccentricity of the hole.  Thus, the first higher-order mode is TE$_{\mathrm{o}11}$ for $e<0.89$, and TE$_{\mathrm{e}21}$ for $e>0.89$. As previously discussed, both even and odd versions of the same mode degenerate into a unique mode as the eccentricity of the ellipse is near zero. 

\begin{figure}[t]
	\subfloat[]{
		\label{TEe11_01}
		\includegraphics[width=0.145\textwidth]{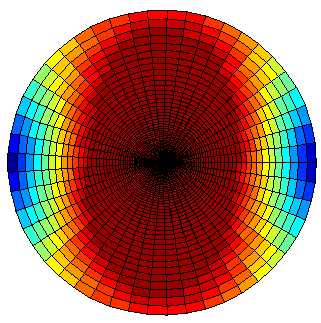}}
	\subfloat[]{
		\label{TEo11_01}
		\includegraphics[width=0.145\textwidth]{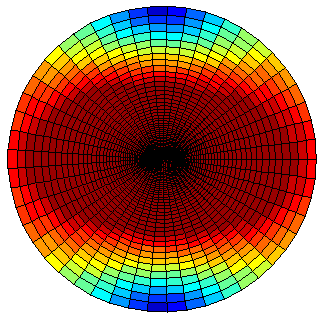}}
	\subfloat[]{
		\label{TMe01_01}
		\includegraphics[width=0.145\textwidth]{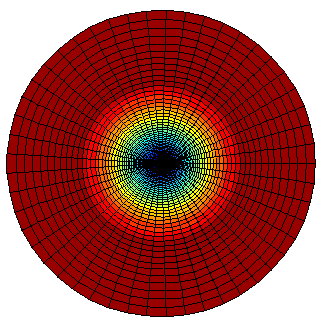}}\\
	\hspace*{1.25cm}
	\subfloat[]{
		\label{TEe21_01}
		\includegraphics[width=0.145\textwidth]{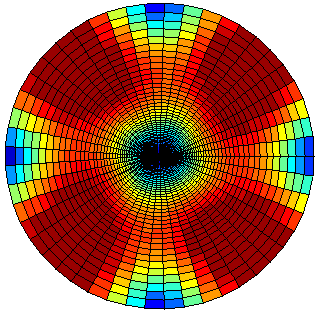}}
	\subfloat[]{
		\label{TEo21_01}
		\includegraphics[width=0.145\textwidth]{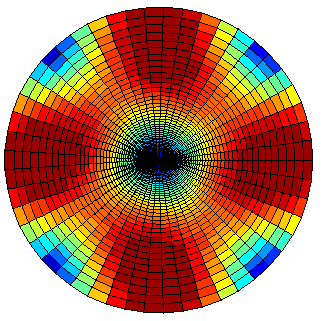}}
	\hspace*{0.35cm}
	\includegraphics[width=0.048\textwidth]{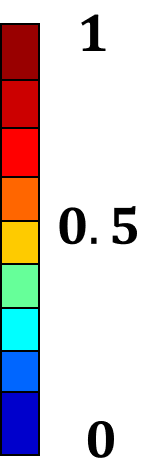}\\
	\caption{\small Normalized absolute value of the electric field distribution of the first five modes inside an elliptical waveguide with eccentricity $e=0.1$: (a) TE$_{\mathrm{e}11}$, (b) TE$_{\mathrm{o}11}$, (c) TM$_{\mathrm{e}01}$, (d) TE$_{\mathrm{e}21}$, (e) TE$_{\mathrm{o}21}$.  }
	\label{plot_modes_01}
\end{figure}
\begin{figure}[!h]
	\subfloat[]{
		\label{TEe11_015}
		\includegraphics[width=0.145\textwidth]{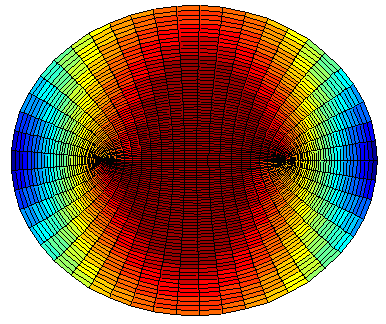}}
	\subfloat[]{
		\label{TEo11_05}
		\includegraphics[width=0.145\textwidth]{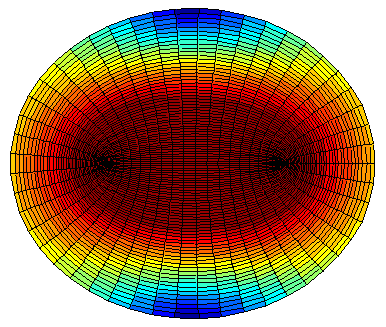}}
	\subfloat[]{
		\label{TMe01_05}
		\includegraphics[width=0.145\textwidth]{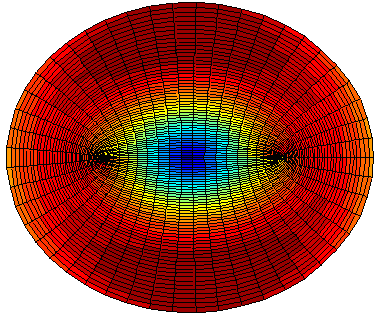}}\\
	\hspace*{1.25cm}
	\subfloat[]{
		\label{TEe21_05}
		\includegraphics[width=0.145\textwidth]{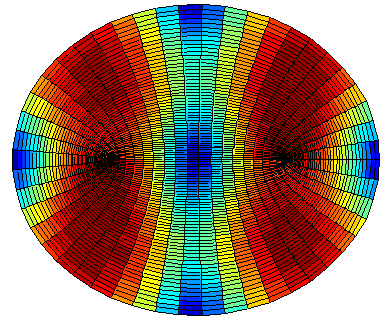}}
	\subfloat[]{
		\label{TEo21_05}
		\includegraphics[width=0.145\textwidth]{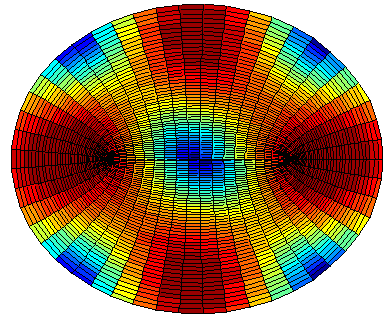}}
	\hspace*{0.35cm}
	\includegraphics[width=0.048\textwidth]{bar_png}\\
	\caption{\small  Normalized absolute value of the electric field distribution of the first five modes inside an elliptical waveguide with eccentricity $e=0.5$: (a) TE$_{\mathrm{e}11}$, (b) TE$_{\mathrm{o}11}$, (c) TM$_{\mathrm{e}01}$, (d) TE$_{\mathrm{e}21}$, (e) TE$_{\mathrm{o}21}$.  }
	\label{plot_modes_05}
\end{figure}


Afterwards, the modal functions are computed using \eqref{modal1}--\eqref{modal4} according to the $q$-parameters derived from Fig. \ref{sort_modes}. The normalized absolute value of the electric field distribution of the first five modes in the elliptic hole (TE$_{\mathrm{e}11}$, TE$_{\mathrm{o}11}$, TM$_{\mathrm{e}01}$, TE$_{\mathrm{e}21}$, TE$_{\mathrm{o}21}$) is shown in Figs. \ref{plot_modes_01} and \ref{plot_modes_05} for two different eccentricities: $e=0.1$ and $e=0.5$. For the case of $e=0.1$ (Fig. \ref{plot_modes_01}), the elliptic waveguide behaves similarly to  a circular waveguide. Thus, the odd modes (Fig. \ref{plot_modes_01}(b) and Fig. \ref{plot_modes_01}(e)) are essentially rotated versions of the even modes (Fig. \ref{plot_modes_01}(a) and Fig. \ref{plot_modes_01}(d)). However, as the eccentricity becomes higher, the even and odd versions of the same mode are not the rotated version of each other (see Figs. \ref{plot_modes_05}(d) and \ref{plot_modes_05}(e)).

\end{appendices}

\ifCLASSOPTIONcaptionsoff
  \newpage
\fi


\end{document}